\documentclass[preprint,12pt]{elsarticle}

\graphicspath{ {./figures/} }
\usepackage{hyperref}
\usepackage{float}
\usepackage{verbatim} %comments
\usepackage{apalike}
\usepackage{amsmath}
\usepackage{amssymb}
\usepackage{multirow}
\usepackage{graphicx}
\usepackage{tabularx}
\usepackage{color}
\usepackage{booktabs}
\usepackage{lscape}
\usepackage{placeins}
\usepackage{caption}
\usepackage{subcaption}
\restylefloat{figure}
\restylefloat{table}

\journal{}

\biboptions{authoryear}

\newcommand{\modelname}{DeepUnifiedMom}
\newcommand{\modelnamefast}{DeepUnifiedMom(Fast)}
\newcommand{\modelnamemedium}{DeepUnifiedMom(Medium)}
\newcommand{\modelnameslow}{DeepUnifiedMom(Slow)}
\newcommand{\modelnamecan}{DeepUnifiedMom(CAN)}
\newcommand{\modelnameeqwt}{DeepUnifiedMom(EQWT)}
\newcommand{\modelnamemvo}{DeepUnifiedMom(MVO)}

\begin{document}

\begin{frontmatter}

% \begin{titlepage}
% \begin{center}
% \vspace*{1cm}

% \textbf{DeepUnifiedMom: Unified Time-series Momentum Portfolio Construction via Multi-Task Learning with Multi-Gate Mixture of Experts}

% \vspace{1.5cm}

% % Author names and affiliations
% Joel Ong$^{a}$ (joel$\_$ong@mymail.sutd.edu.sg), Dorien Herremans$^a$ (dorien$\_$herremans@sutd.edu.sg) \\

% \hspace{10pt}

% \begin{flushleft}
% \small  
% $^a$ Singapore University of Technology and Design, 8 Somapah Rd, Singapore 487372 \\

% \begin{comment}
% Clearly indicate who will handle correspondence at all stages of refereeing and publication, also post-publication. Ensure that phone numbers (with country and area code) are provided in addition to the e-mail address and the complete postal address. Contact details must be kept up to date by the corresponding author.
% \end{comment}

% \vspace{1cm}
% \textbf{Corresponding Author:} \\
% Joel Ong \\
% Singapore University of Technology and Design, 8 Somapah Rd, Singapore 487372 \\
% Tel: (65) 96951117 \\
% Email: joel$\_$ong@mymail.sutd.edu.sg

% \end{flushleft}        
% \end{center}
% \end{titlepage}


% \includegraphics[width=1\textwidth]{graphical_abstract_mom_mmoe.jpg}

% % Research highlights
% \begin{highlights}
% \item Model for unified momentum portfolios across multiple time frames.
% \item First use of deep multi-gate mixture of expert multi-task learning in portfolios.
% \item Extensive analysis on performance against various benchmarks.
% \end{highlights}

\title{DeepUnifiedMom: Unified Time-series Momentum Portfolio Construction via Multi-Task Learning with Multi-Gate Mixture of Experts}

\author[label1]{Joel Ong}
\ead{joel\_ong@mymail.sutd.edu.sg}

\author[label1]{Dorien Herremans}
\ead{dorien\_herremans@sutd.edu.sg}

\cortext[cor1]{Joel Ong}
\address[label1]{Singapore University of Technology and Design, 8 Somapah Rd, Singapore 487372}

\begin{abstract}

This paper introduces \textsf{\modelname}, a deep learning framework that enhances portfolio management through a multi-task learning approach and a multi-gate mixture of experts. The essence of \textsf{\modelname} lies in its ability to create unified momentum portfolios that incorporate the dynamics of time series momentum across a spectrum of time frames—a feature often missing in traditional momentum strategies. Our comprehensive backtesting, encompassing diverse asset classes such as equity indexes, fixed income, foreign exchange, and commodities, demonstrates that \textsf{\modelname} consistently outperforms benchmark models, even after factoring in transaction costs. This superior performance underscores \textsf{\modelname}'s capability to capture the full spectrum of momentum opportunities within financial markets. The findings highlight \textsf{\modelname} as an effective tool for practitioners looking to exploit the entire range of momentum opportunities. It offers a compelling solution for improving risk-adjusted returns and is a valuable strategy for navigating the complexities of portfolio management.
\end{abstract}


\end{frontmatter}

% \tableofcontents %just for ease of reviewing

\section{Introduction}
\label{introduction}

Time-series momentum (TSMOM) strategies are a systematic approach in finance that leverages the persistence of asset returns over time. These strategies aim to exploit the continuation of underlying trends by establishing long positions during uptrends and short positions during downtrends~\citep{jegadeesh_titman_1993,jegadeesh_titman_2001}. The concept of momentum has garnered extensive attention in financial literature, underscoring its importance. Research by ~\citet{moskowitz_2012} highlights the effectiveness of TSMOM, showcasing impressive risk-adjusted returns by simply buying assets with positive past 12-month returns. Further studies across various asset classes corroborate these findings, emphasizing the robustness of TSMOM strategies~\cite{georgopoulou_wang_2016, levine_2016, hurst_2017, yulia_2020}. At the core of TSMOM strategies is volatility scaling, a crucial method for managing exposure to market volatility~\citep{baltas_2012}. Through adjustments in exposure levels during periods of low and high volatility, TSMOM strategies effectively mitigate the risk of significant losses during market turbulence~\citep{cambell_2020}. This approach has proven invaluable in enhancing Sharpe ratios, curbing extreme tail returns, and limiting maximum drawdowns in portfolios of risky assets~\citep{barroso_2015, daniel_2016, ong_2023}.

However, TSMOM strategies often fail to account for the interactions between different assets within a portfolio~\citep{nilsson_2015, xingyue_2023}. This oversight can lead to excessive risk exposure, as these strategies treat each asset in isolation without considering how they correlate and interact. For example, the simultaneous momentum trends in equities, commodities, and currencies can amplify overall portfolio risk if not managed cohesively. Overlooking these interactions can diminish diversification benefits and increase the likelihood of significant drawdowns during market turbulence.

Moreover, various asset classes, and even individual assets within those classes, exhibit distinct momentum dynamics with differing trend speeds. This variation makes risk allocation challenging, as applying a one-size-fits-all approach to trend speed can result in less-than-ideal investment outcomes~\citep{levine_2016, duan_2023}. To address this, some practitioners implement multiple TSMOM portfolios, each tailored to a specific trend speed, and distribute capital among them~\citep{tzotchev_2018, zambrano_2022, goulding_2023}. Nevertheless, this approach can still result in inefficient capital distribution, indicating a need for more refined methods that can adeptly navigate the varying momentum speeds of different assets and asset classes for more effective capital allocation.

We propose an innovative approach to bridge the research gaps identified earlier inspired by recent advancements in deep learning for portfolio construction. The proposed method utilizes a deep Multi-task Learning framework combined with a Multi-gate Mixture-of-Experts architecture to develop a momentum portfolio~\citep{jacobs_1991, ma_2018}. Since its development more than three decades ago, the Mixture-of-Experts (MoE) approach has become foundational in numerous research areas and has recently been pivotal in advancing the field of natural language processing in large language models (LLMs)~\citep{fedus_2022, zoph_2022, he_jiaao_2022, gale_2022, sheng_2023}. Our approach, which we call \textsf{\modelname}, aims to seamlessly integrate momentum opportunities across various speeds, enhancing the efficacy of traditional momentum strategies. Our main contributions are: i) introducing a novel Multi-task Learning framework with a Multi-gate Mixture-of-Experts architecture, which facilitates end-to-end learning for multi period portfolio construction to enhance momentum portfolio performance. ii) This study represents the first implementation and examination of Multi-task Learning combined with a Multi-gate Mixture-of-Experts approach, specifically within portfolio construction. iii) We provide a comprehensive experimental analysis to evaluate and understand the performance outcomes of this innovative methodology against existing momentum strategies and various portfolio construction techniques.

The paper is organized as follows. In Section \ref{sec:related_work}, we discuss existing work on contructing both classical and deep-learning momentum portfolios. Section \ref{sec:methodology} presents the proposed \textsf{\modelname} model. Next, in Section \ref{sec:expertimental_setup} we present the setup of the different experiments, including a description of the dataset, benchmark models, and proposed backtesting strategy. In Section~\ref{sec:performance_evaluation}, the results of our experiments are presented. Finally, in Section~\ref{sec:conclusion} we summarize our findings and suggest directions for future research.

\section{Related Work}
\label{sec:related_work}

Research in deep learning in finance is well-established, with numerous studies leveraging these techniques to enhance prediction accuracy, portfolio optimization, and risk assessment. \cite{zhao_2023} introduce a hybrid model, SA-DLSTM, combining emotion-enhanced convolutional neural networks (ECNN), denoising autoencoders (DAE), and long short-term memory (LSTM) models to predict stock price movements by analyzing sentiment from user-generated comments. \cite{wang_2022} propose a portfolio construction model integrating the KMV model with a multiobjective water cycle algorithm, enhancing portfolio evaluation and stability by incorporating financial data from listed companies. \cite{min_2021} address conservatism in worst-case robust portfolio optimization by suggesting hybrid models that use LSTM and XGBoost to forecast market movements and generate hyperparameters for modeling. \cite{lin_2022} develop a multiagent-based deep reinforcement learning framework for portfolio management, featuring a two-level nested agent structure and a custom reward function to optimize trading decisions and risk transfer behaviors. Lastly, \cite{ozbayoglu_2020} provide a comprehensive survey of deep learning applications in finance, categorizing models and identifying future research opportunities. These studies collectively demonstrate advancements in applying deep learning to financial applications, highlighting the potential for innovative techniques to improve financial model robustness and performance. However, none of these studies specifically focus on momentum portfolio construction.

\cite{moskowitz_2012} introduced the concept of time-series momentum (TSMOM), demonstrating that the excess returns of an asset over the past 12 months strongly predict its future performance. Since introducing this concept, it has become a conventional practice for practitioners to implement momentum-based portfolios~\citep{asness_2014, hurst_2017, baltas_2021}. In recent years, the use of deep learning approaches in portfolio construction has gained increasing popularity~\citep{bryan_2019, zhang_zohren_roberts_2020, yanzhe_2022, yu_2022, xingyue_2023, chenxun_2023}. More notably, in the space of deep learning for portfolio construction, ~\cite{wood_giegerich_roberts_zohren_2022} proposed deep-learning architecture that improves upon traditional time-series momentum and mean-reversion strategies. With multiple attention heads~\citep{vaswani_2017}, it tracks diverse market regimes across timescales and offers interpretability by highlighting influential factors and key time steps, refining trading strategies. The attention mechanism helps to enhance learning of long-term dependencies and adaptability to new market conditions like the SARS-CoV-2 crisis.~\cite{ong_2023} first proposed the application of deep multi-task learning (MTL) in momentum portfolio construction, which incorporates auxiliary tasks related explicitly to volatility forecasting~\citep{parkinson_1980, garman_1980, rogers_1991, yang_2000}. The findings highlighted that a comprehensive MTL framework, encompassing all proposed auxiliary tasks, not only enhances the risk-adjusted performance of portfolios but also notably reduces maximum drawdowns compared to deep learning models without auxiliary tasks or those with a single auxiliary task. This approach underscores the critical role of effectively selecting auxiliary tasks in MTL settings to optimize portfolio outcomes.

Despite extensive research in the field, a significant gap persists in the development of a unified momentum portfolio capable of adapting to trends of varying speeds. Previous studies have typically focused independently on fast (less than one month), medium (three to six months), or slow (six months to one year) momentum strategies, rather than integrating these approaches into a single, dynamic framework. A unified momentum approach aims to capture momentum across this spectrum, seamlessly adjusting to the changing pace of trends within assets and asset classes in the portfolio. Addressing this gap in the literature represents a crucial advancement that could significantly enhance our understanding of momentum-based trading and lead to more robust portfolio management strategies. By developing such a portfolio, we can better exploit the full range of momentum opportunities in financial markets, resulting in superior risk-adjusted returns compared to existing momentum strategies in the literature.

This work aims to bridge this research gap by presenting a deep learning approach to constructing a unified momentum portfolio. We begin by applying the principles of multi-task learning to train three task-specific networks, each specializing in predicting the forward momentum signal score for one month, three months, and six months ahead. These networks generate portfolios representing fast, medium, and slow momentum strategies. The outputs from these task-specific networks are then fed into a final network, the Capital Allocation Module, which determines the weight allocation for each portfolio. By distributing risk according to the allocations provided by the Capital Allocation Module, we create a unified momentum portfolio that effectively exploits and leverages both short-term and long-term trends across assets and asset classes in the portfolio.

\section{Methodology}
\label{sec:methodology}

\subsection{Overview}

This work proposes to include Multi-Gate Mixture of Experts with a Multi-Task Learning Architecture~\citep{ma_2018}. This novel approach combines Long-Short Term Memory (LSTM) ~\citep{schmidhuber_1997} modules with the Multi-Gate Mixture of Experts (MoMME) framework~\citep{jacobs_1991} to construct a unified momentum portfolio.

\begin{figure}[ht!]
  \centering
  \includegraphics[width=0.9\linewidth]{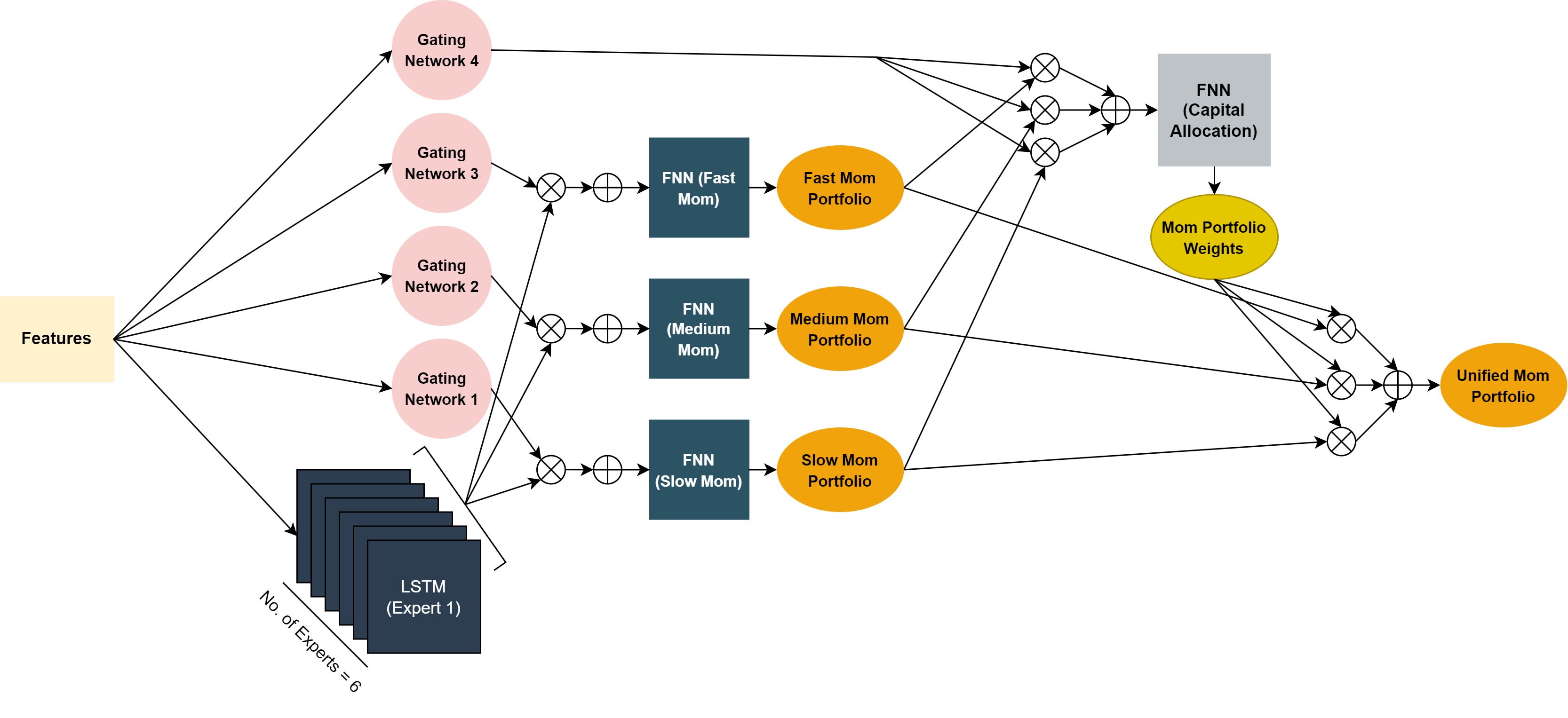}
  \caption{The figure illustrates the proposed architecture, highlighting the flow from shared LSTM experts through task-specific gating and FNN layers and culminating in a final FNN that determines portfolio weights. These weights allocate risk across the portfolios generated by the task-specific FNN layers, which include three momentum portfolios (Fast, Mid, and Slow), each tailored to different trend speeds. The overarching goal of the final unified momentum portfolio is to capitalize on diverse market trends strategically. The symbol $\bigotimes$ followed by $\bigoplus$ denotes the weighted sum of the outputs by the gating network with either the LSTM experts or the FNN task-specific network output.}
  \label{fig:mom_mmoe.drawio.png}
\end{figure}

In our proposed architecture, the LSTM experts serve as shared layers forming the backbone of our multi-task learning framework. This setup enables effective parameter sharing across various task-specific pathways yielding two key benefits~\citep{caruana_1997, thrun_1998, sebastian_2017}. Firstly, the shared LSTM experts facilitate a more efficient learning process by leveraging commonalities among tasks and consolidating learning efforts. This approach accelerates the training process and enhances overall model performance by drawing on a broader base of data insights. Secondly, the use of shared experts enhances the model's generalization capabilities. By exposing the model to a variety of tasks within the same learning process, it becomes less prone to overfitting on any single task~\citep{ghosn_1996, baxter_2000, sebastian_2017, lukas_2018}. Overall, the integration of shared LSTM experts within our architecture underscores a strategic approach to harnessing the complexities of financial data.

Each task-specific network has a dedicated gating network, a cornerstone of the Multi-Gate Mixture of Experts (MoMME) framework~\citep{ma_2018}. In our work, these gating networks are specialized one-layer FNNs with a softmax activation function. They receive the same feature inputs as the LSTM experts and output a set of weights that sum to one. These weights determine the reliance on the corresponding LSTM experts. By selectively activating relevant LSTM experts, the gating networks enhance the performance of task-specific networks. This selective activation enables more effective learning, as each task-specific network can focus on constructing momentum portfolios tailored to specific speeds or timeframes, ultimately improving the performance of the constructed momentum portfolios. Here, three task-specific networks are trained to minimize the root mean square error (RMSE) between their outputs and the forward-looking TSMOM signals with one-month, three-month, and six-month lookback timeframes. The outputs of these task-specific networks yield the fast, medium, and slow momentum portfolios. Finally, we have a task-specific network called the Capital Allocation Network, supported by a gating network. The gating network receives feature inputs and assigns appropriate weights to the outputs of each task-specific network. The weighted outputs are then fed into the Capital Allocation Network, which is trained to allocate weights to the fast, medium, and slow momentum portfolios generated by the three preceding task-specific networks. This results in a final unified momentum portfolio that strategically capitalizes on opportunities across various market trends.

The output produced by the Capital Allocation Network serves as a set of weights assigned to the fast, medium, and slow momentum portfolios generated by the preceding task-specific networks. These weights determine the allocation of capital across the different portfolios, reflecting the model's strategic decisions on how to distribute resources among various market trends. By optimizing these weights, the Capital Allocation Network aims to construct a final unified momentum portfolio that effectively captures opportunities across diverse market conditions. Essentially, the portfolio weights determined by the Capital Allocation Network represent the model's assessment of the relative importance and potential profitability of each momentum portfolio. In summary, the output of the Capital Allocation Network, in conjunction with the portfolio weights for the fast, medium, and slow momentum portfolios, collectively yield the final unified momentum portfolio. This integrated approach enables the model to adaptively allocate resources and strategically capitalize on market trends, ultimately enhancing portfolio performance.

To sum up, our proposed deep mixture of experts' multi-gate and multi-task learning architecture generates three distinct momentum portfolios, each designed to capture momentum at different timeframes. The Capital Allocation Network also allocates weights to these three momentum portfolios, constructing the final unified momentum portfolio. This approach enables the creation of a unified momentum portfolio in a single step, with the model optimized in an end-to-end fashion. A single-step, end-to-end optimized model is superior because it ensures that all components are trained simultaneously, allowing for seamless integration and interaction between different parts of the model. This approach enhances overall performance by effectively capturing dependencies and relationships within the data, resulting in a more cohesive and robust final portfolio. Our experimental results, detailed in Section~\ref{sec:performance_evaluation}, substantiate this claim.

\subsection{Multi-Task Learning Network}

By categorizing momentum into fast (one-month), medium (three-month), and slow (six-month) categories, each task-specific network within our Multi-Task Learning framework is trained to construct portfolios that capture momentum at their respective time frame. This segmentation enhances the model's ability to detect and leverage the subtle variations in momentum within each asset class and individual asset, thereby improving the precision and relevance of its predictions for future momentum returns across diverse assets. Each task-specific network is represented by a Feedforward Neural Network (FNN), trained to predict the forward-looking time-series momentum (TSMOM) signal. The TSMOM signal is essentially the forward return of an asset, adjusted for its volatility. This adjustment takes into account the risk associated with the asset, providing a normalized measure of momentum that is more comparable across different assets. During the training process, the objective is to minimize the difference between the predicted TSMOM signal and the actual forward-looking TSMOM signal (the ground truth, denoted as $\hat{y}^{i}_{t}$). Specifically, we minimize the Root Mean Squared Error (RMSE) between the predicted output $y^{i}_{t}$ and the ground truth $\hat{y}^{i}_{t}$. The RMSE is a commonly used metric for regression tasks, providing a measure of the average magnitude of the errors between predicted and actual values.

\begin{equation} \label{eq:1}
L_{\text{RMSE}} = \frac{1}{B} \sum_{t \in B} l_{\text{RMSE}_{t}}
\end{equation}

where,

\begin{align*}
l_{\text{RMSE}_{t}} = \sqrt{\frac{1}{n} \sum_{i=1}^{n} (\hat{y}^{i}_{t} - y^{i}_{t})^2}
\end{align*}

\noindent where $L_{\text{RMSE}}$ represents the overall RMSE loss over a batch of size $B$. The term $l_{\text{RMSE}_{t}}$ denotes the RMSE at specific time $t$ in the batch. In this context, $n$ is the total number of assets in the portfolio and $y^{i}_{t}$ is the prediction output by the task-specific network for the forward-looking TSMOM signal for asset $i$ at time $t$. Moreover, the forward-looking TSMOM signal for each asset $i$ at time $t$, denoted as $\hat{y}_{i}$ represents the ground truth, or the actual observed value of the forward-looking TSMOM signal for each asset $i$ at time $t$. This ground truth is used during the training phase of your model to compare against the predicted output $y^{i}_{t}$. It is defined as:

\begin{equation}\label{eq:2}
\hat{y}^{i}_{t} = \text{TSMOM}^{i}_{t} = \frac{r_{t+1, t+s}^{i}}{\sigma_{t+s}^{i}}
\end{equation}

In this formulation, $r_{t+1, t+s}^{i}$ represents the returns from $t + 1$ to $t + s$, and $\sigma_{t+s}^{i}$ is the standard deviation of the returns, both calculated at a future window $s$ beyond time $t$. Here, the window $s$ varies according to the speed category of the momentum being analyzed: 20 trading days for \textsf{\modelnamefast}, 60 trading days for \textsf{\modelnamemedium}, and 120 trading days for \textsf{\modelnameslow}. This variation allows each task-specific network to fine-tune its learning process to the particular momentum time frame it addresses, thus enhancing the model's ability to adapt to the distinct market dynamics associated with each speed category. Finally, the return of the fast, medium and slow portfolio can be calculated as follows:

\begin{equation} \label{eq:3}
r_{t,t+1}^{\rho} = \frac{1}{n} \sum_{i=1}^{n}  y_{t-1, t}^{\rho, i} \times \; r_{t,t+1}^{i}
\end{equation}

\noindent where $\rho$ represents the \textsf{\modelname}-Fast, Medium and Slow portfolio, $n$ is the number of assets in the portfolio, $y_{t-1, t}^{\rho, i}$ is the output of the task-specific network (indicating the weight allocated to asset $i$ for the given momentum timeframe) at time~$t$, $r_{t, t+ 1}^{i}$ is the one day return of the asset $i$ and $r_{t,t+1}^{\rho}$ is the $\rho$ portfolio's return at time $t$.

\subsection{Capital Allocation Network}

The objective of the Capital Allocation Network (CAN) is to generate weights for allocating capital across the various momentum portfolios produced by the task-specific networks. It is implemented as a specialized feedforward neural network (FNN) with a tanh activation function at each intermediate layer and a softmax activation function at the final layer to ensure the output sums to 1. By applying the weights generated by the CAN to \textsf{\modelname}-Fast, Medium and Slow, we obtain the final unified momentum portfolio, which we term \textsf{\modelnamecan}. The unified momentum portfolio's return can be calculated as follows:

\begin{equation} \label{eq:4}
r_{t,t+1}^{\cup} = \sum_{\rho \in P}  w_{t-1,t}^{\rho} \times \; r_{t,t+1}^{\rho}
\end{equation}

\noindent where $P$ is the set of fast, medium and slow momentum portfolios, $w_{t-1,t}^{\rho}$ is the weight predicted by the CAN for portfolio $\rho$ at time~$t$, $r_{t, t+ 1}^{i}$ is the one day return of the asset $i$ and $r_{t,t+1}^{\cup}$ is the return of the unified momentum portfolio.

Since the objective of the CAN differs from those of the task-specific networks, we use the Sharpe Ratio~\citep{sharpe_1994} as the objective function for training. By utilizing the Sharpe Ratio, as outlined in Equation \ref{eq:5}, we direct the model to learn how to generate portfolios optimized for risk-adjusted returns from the input features. The Sharpe Ratio measures an investment's performance relative to a risk-free asset, adjusting for risk, and offers a comprehensive metric for evaluating the trade-off between risk and return. Incorporating this ratio into the model's learning process ensures that the constructed portfolios aim to maximize returns while minimizing risk, resulting in superior risk-adjusted performance~\citep{bryan_2019, zhang_zohren_roberts_2020, ong_2023}.

\begin{figure}[ht!]
  \centering
  \includegraphics[width=0.9\linewidth]{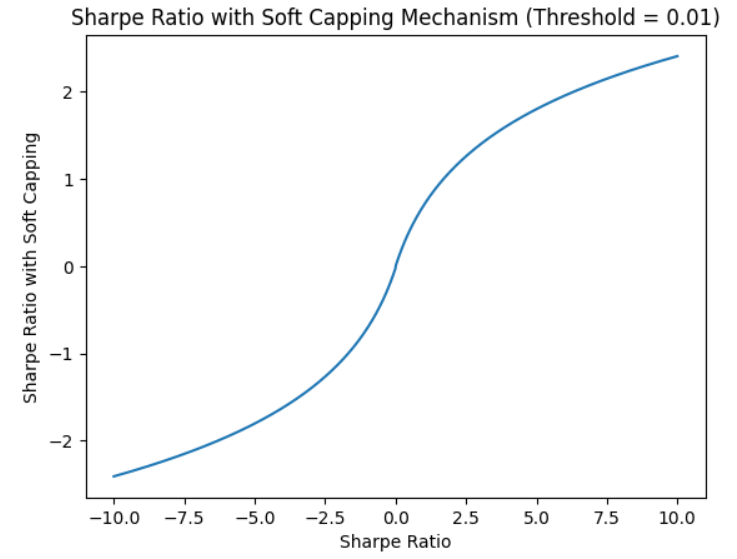}
  \caption{Relationship between the Sharpe Ratio with Soft Capping Mechanism and the Standard Sharpe Ratio with Threshold = 0.01}
  \label{fig:sharpe_ratio}
\end{figure}

\begin{equation} \label{eq:5}
L_{\text{Sharpe Ratio}^{\cup}} = -\frac{\mathbb{E}[r^{\cup}]}{\sigma_{r^{\cup}}}
\end{equation}

\noindent where $\mathbb{E}[r^{\cup}]$ and $\sigma_{r^{\cup}}$  are the mean and standard deviation of the unified momentum portfolio's realised returns, respectively. The high noise-to-signal ratio in financial data significantly increases the risk of overfitting in deep learning models~\citep{vitells2011estimating, harvey2014evaluating, bailey2015probability, lopez2018advances, israel2020can}. During training, we may encounter instances where certain batches have an extremely high noise-to-signal ratio. Fitting to this noise can result in a high Sharpe Ratio. Consequently, the model, aiming to maximize the Sharpe ratio, may overfit to these noisy patterns, allocating more weight to these instances and performing well on the training data but failing to generalize to new data. We introduce a modified objective function called the Sharpe Ratio with a Soft Capping Mechanism to mitigate this risk. It first caps the Sharpe ratio at a specified threshold value, ensuring that any value above this threshold is limited. For values that exceed this threshold, the function computes the excess and applies a logarithmic transformation, which reduces the impact of these extreme values by making them grow more slowly. Similarly, it ensures that the Sharpe ratio does not fall below the negative of this threshold by capping the lower end and applying a logarithmic transformation to values below the threshold. This combination of capping and logarithmic adjustments smooths out the extremes, as shown in the equation below:

\begin{equation} \label{eq:6}
L_{\text{SR}_{\text{soft}}} = -(L + \log(1 + U_e) - \log(1 - L_e))
\end{equation}

where,

\begin{align*}
U &= \min(\text{SR}, \tau) \\
U_e &= \max(\text{SR} - \tau, 0) \\
L &= \max(U, -\tau) \\
L_e &= \min(\text{SR} - \tau, 0)
\end{align*}

Here, $\text{SR}_{soft}$ represents the modified Sharpe ratio with soft capping mechanism, $\text{SR}$ represents the original Sharpe ratio, and $\tau$ is the threshold, which is set to 0.01 during the training process. The resulting Sharpe Ratio with Soft Capping mechanism can be seen in Figure~\ref{fig:sharpe_ratio}. The logarithmic transformation applied to values exceeding the threshold moderates their growth, reducing the impact of extreme values. This smoothing effect stabilizes the training process by preventing abrupt changes in the model's behavior due to outliers. Consequently, the model is encouraged to focus on more consistent and reliable patterns in the data, leading to better generalization. As a result, the model is less likely to overfit to noise and more likely to capture true underlying signals that are relevant to the objective at hand. Our experimental results, detailed in Section~\ref{sec:expertimental_setup} will substantiate the effectiveness of training our proposed architecture with the Sharpe Ratio with a Soft Capping Mechanism. This comparison with models trained without the mechanism will highlight the improvement of performance and generalization capabilities when using the Sharpe Ratio with the Soft Capping Mechanism.

\subsection{Loss function}

\begin{equation}  \label{eq:7}
\begin{split}
L_{total} &= L_{\text{SR}_{\text{soft}}} + \sum_{\rho \in P} \; L_{\text{RMSE}}^{\rho}\\
\end{split}
\end{equation}

Putting it all together, the final loss function of our model, which we minimize during training, can be written as Equation~(\ref{eq:7}). In this equation, $L_{total}$ combines two components. The first component, $L_{\text{SR}_{\text{soft}}}$, represents the loss for the CAN, calculated as the negative Sharpe ratio with a soft capping mechanism. The second component is the sum of the RMSE losses $L_{\text{RMSE}}^{\rho}$ for each task-specific network. The overall loss function integrates these elements to guide the training process and optimize the model's performance.

\section{Experimental Setup}
\label{sec:expertimental_setup}

\subsection{Dataset}
\label{sec:dataset} 

Individual futures contracts are subject to expiration dates and varying levels of liquidity, which can hinder the practical analysis of long-term trends. To overcome this challenge, we rely on the Pinnacle Data Corp CLC database as our primary data source for evaluating the proposed model. The data we used in our experimentation spans from January 1990 to December 2023, providing daily data and encompassing over three decades of historical information. The Pinnacle Data Corp CLC database offers a comprehensive continuous price history for 49 futures contracts across diverse asset classes, such as commodities, currencies, fixed income, and equity index futures. We leverage the continuous contract history of each asset, constructed through end-to-end concatenation and price adjustment using the backward-ratio method, to ensure robust analysis of long-term trends.

\subsection{Feature Set}
\label{sec:feature_set}
We derive a set of time-series momentum features from the daily settled price of the continuous futures by taking the log returns ($r_{t - d, t}^{i}$) over the past 3 trading days, 5 trading days, 10 trading days, 21 trading days, 63 trading days, 126 trading days, and finally 252 trading days: 

\begin{equation} \label{eq:8}
r_{t - d, t}^{i} = \ln{\frac{P_t^i}{P_{t - d}^i}}
\end{equation}

\noindent where $r_{t - d, t}^{i}$ is the natural logarithm of the $d$-day return of the asset $i$ at day $t$, $P_{t}^{i}$ is the settled price of asset $i$ at time $t$ and $P_{t-d}^{i}$ is the settled price of asset $i$, $d$ trading days ago at time $t$. To normalize the returns and account for the variability in market conditions, we scale the calculated log returns by the asset's volatility. This approach ensures that the returns are standardized, allowing for a more equitable comparison across different assets and time periods. The scaled return, $\hat{r}_{t - d, t}^{i}$, is computed as follows:

\begin{equation} \label{eq:9}
\hat{r}_{t - d, t}^{i} = \frac{r_{t - d, t}^{i}}{\sigma_{t-d, t}^{i}}
\end{equation}

\noindent where $\hat{r}_{t - d, t}^{i}$ represents the volatility-normalized return over the $d$-day period for asset $i$ at day $t$, $r_{t - d, t}^{i}$ is the log return as previously defined, and $\sigma_{t-d, t}^{i}$ denotes the volatility of the asset over the $d$-day period.

Following the approach of ~\citet{ong_2023}, our methodology for feature creation is guided by two principal considerations. Firstly, we aim to preserve the essence of time-series momentum by utilizing features that align closely with those employed in the construction of time-series momentum portfolios, as outlined by ~\citet{moskowitz_2012}. Secondly, and of greater significance, we intentionally limit the complexity of our feature engineering to ensure that the observed performance of the portfolios is primarily attributed to the efficacy of our architectural design, rather than to the ingenuity or specificity of the features used. This approach underscores our commitment to validating the inherent strength and adaptability of the architecture in capturing momentum trends, rather than leveraging elaborate feature engineering to enhance portfolio performance artificially.

\subsection{Benchmark Models}

The concept of Time-Series Momentum (TSMOM) portfolios, as introduced by \citet{moskowitz_2012}, forms the cornerstone of our benchmarking process. These portfolios operate on the principle of buying or selling assets based on their performance over the past 12 months. To comprehensively assess our model's efficacy, we have meticulously crafted a suite of TSMOM portfolios, each tailored to capture distinct momentum horizons:

\begin{itemize}
    \item TSMOM(1): based on the past one month's returns.
    \item TSMOM(3): based on the past three month's returns.
    \item TSMOM(6): based on the past six month's returns.
    \item TSMOM(12): based on the past twelve month's returns.
    \item TSMOM(1,4): An equal-weighted combination of the 1, 2, 3, and 4-month TSMOMs.
    \item TSMOM(5,8): An equal-weighted combination of the 5, 6, 7, and 8-month TSMOMs.
    \item TSMOM(9,12): An equal-weighted combination of the 9, 10, 11, and 12-month TSMOMs.
    \item TSMOM(1,12): An equal-weighted combination of the 1 to 12-month TSMOMs.
\end{itemize}

This wide array of TSMOM portfolios serves as a benchmark, enabling us to evaluate our proposed model against a spectrum of momentum-based investment strategies across different timeframes. To support our claim that a unified momentum portfolio, which can capitalize on momentum opportunities across various timeframes, \textsf{\modelnamecan} should outperform all TSMOM benchmark strategies. In addition to that, to rigorously evaluate the final unified momentum portfolio's performance constructed by the \textsf{\modelnamecan}, we established two benchmark portfolios: \textsf{\modelnameeqwt} and \textsf{\modelnamemvo}. These carefully chosen benchmarks will help us assess the effectiveness of the unified momentum portfolio constructed by the CAN compared to existing standard portfolio construction techniques. 

\begin{itemize}
    \item \textsf{\modelnameeqwt}: This portfolio equally distributes weights from portfolios constructed by \textsf{\modelname}-Fast, Medium and Slow.
    \item \textsf{\modelnamemvo} utilizes Mean-Variance Optimization (MVO) \citep{harry_1952} to construct a portfolio that maximizes the Sharpe ratio.
\end{itemize}

Constructing a portfolio with equal weighting is a straightforward heuristic that does not involve any optimization process. In contrast, constructing a final portfolio using MVO involves a second optimization process by using the historical returns of \textsf{\modelname}-Fast, Medium and Slow portfolios to calculate the expected returns and covariance matrix, which are then optimized by maximizing the Sharpe ratio to construct the final portfolio. The key drawback of such approaches is that they do not consider the interactions between the components holistically. This fragmented optimization can lead to suboptimal overall performance because each step is optimized in isolation without accounting for all components' interdependencies and joint effects. Finally, to substantiate our earlier claim that our proposed model's single-step, end-to-end optimization of a final unified momentum portfolio is more optimal, the final portfolio constructed by the \textsf{\modelnamecan} should outperform the portfolios constructed using both the \textsf{\modelnameeqwt} and \textsf{\modelnamemvo} methods.

\subsection{Backtest Specifications}

\begin{figure}[ht!]
  \centering
  \includegraphics[width=0.9\linewidth]{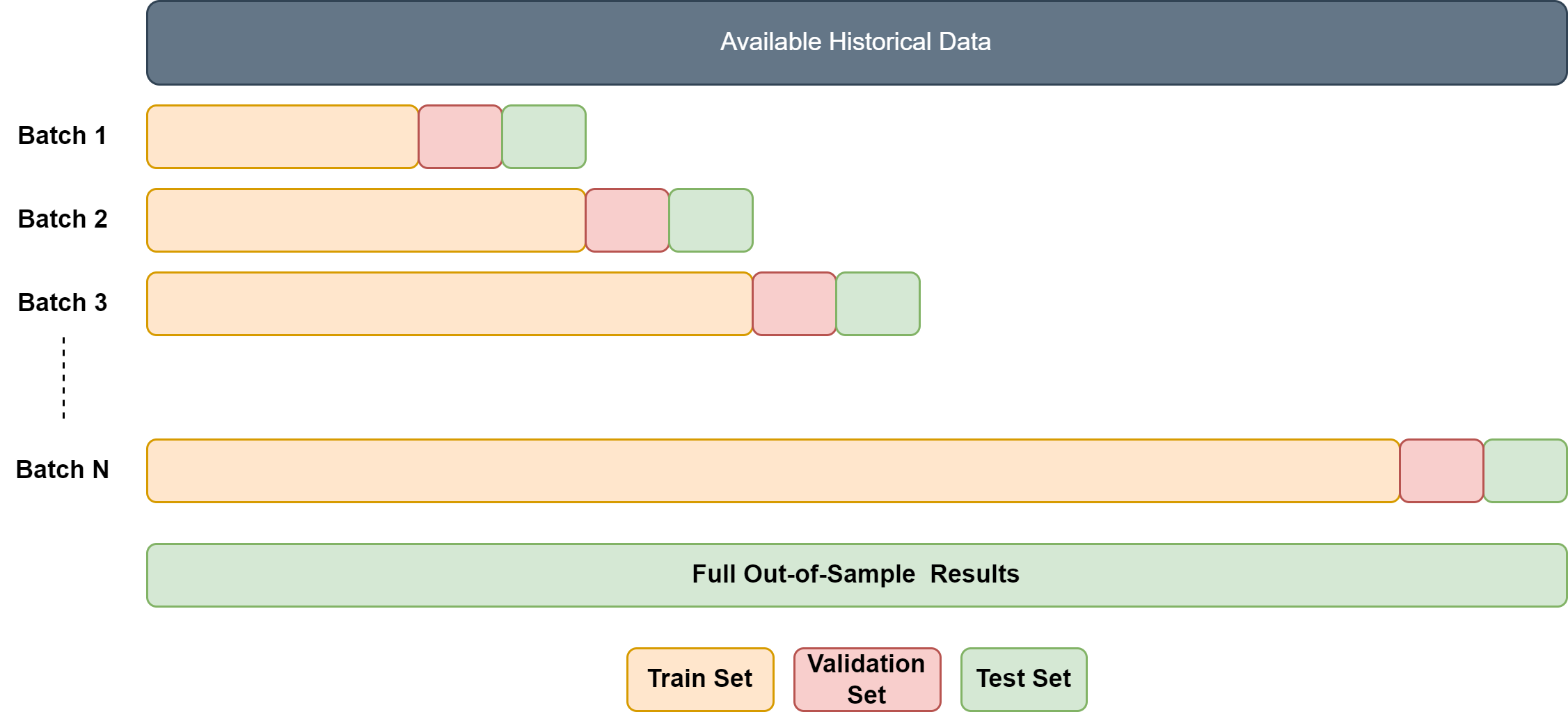}
  \caption{The diagram illustrates the expanding window cross-validation approach used in our model's training.}
  \label{fig:backtest_expanding_window.drawio.png}
\end{figure}

In the following section, we present the backtest results obtained from our proposed model, trained using an expanding window cross-validation approach with the first batch of training data spanning a period of 10 years, as illustrated in Figure~\ref{fig:backtest_expanding_window.drawio.png}. Throughout the training phase, 20\% of the data was allocated for validation purposes. The trained models were then utilized to construct portfolios for the test set, with each portfolio corresponding to a year's worth of out-of-sample data.  This process was repeated 24 times, resulting in an out-of-sample backtest period from January 2000 to December 2023.

\begin{table}[htb]
% \resizebox{\textwidth}{!}{%
% \begin{tabularx}{\linewidth}{lr}
\centering
\caption{Hyperparameter Search Space.}
\begin{tabular}{lr}
\toprule

Parameters & Values\\
\midrule
Number of LSTM Layers & 1, 2, 3\\ 

LSTM Hidden Units &  64, 126, 252, 512 \\ 

Number of LSTM Experts & 3, 6, 9, 12 \\

Task-Specific Netowrk Layers & 2, 3, 4 \\

Task-Specific Network Hidden Units &  64, 126, 252, 512 \\ 

\bottomrule
\end{tabular}
\label{table:1}
% }
\end{table}

During the backtesting phase, our model underwent training on the designated training dataset. We employed Stochastic Gradient Descent (SGD) with the Adam optimizer to minimize the loss functions. We conducted a grid search on the validation set, guided by the parameter search space outlined in Table \ref{table:1}. The training process was designed to conclude after 20 epochs; however, we incorporated an early stopping mechanism that halts training if there's no improvement in validation loss over 5 consecutive epochs. Utilizing an expanding window approach for out-of-sample training and validation, the final training iteration—covering data from January 1990 to December 2023—required approximately 1 hour on a system equipped with an NVIDIA GeForce RTX 2090.

\section{Performance Evaluation}
\label{sec:performance_evaluation}

As shown in Table \ref{table:2}, our final unified momentum portfolio generated by \textsf{\modelnamecan}, outperforms all TSMOM benchmark models, including \textsf{\modelname}-Fast, Medium, and Slow, as well as the final portfolios generated by \textsf{\modelname} - EQWT and MVO. \textsf{\modelnamecan} achieves a Sharpe ratio of 2.33 and a Sortino ratio of 3.88 while incurring a maximum drawdown of -1.02\%. In comparison, the best TSMOM benchmark strategy, TSMOM(1,12), achieves a Sharpe ratio of 1.07 and a Sortino ratio of 1.58, with a drawdown of -2.01\%. This demonstrates the superior performance of \textsf{\modelnamecan} in terms of both risk-adjusted returns and drawdown management. 

Additionally, when comparing the portfolios generated by the task-specific networks \textsf{\modelname}-Fast, Medium, and Slow, to the final unified momentum portfolio, the latter consistently outperforms the former on a risk-adjusted basis, notably achieving a much lower maximum drawdown. The best-performing task-specific network, \textsf{\modelnameslow}, achieves a Sharpe ratio of 1.54 and a Sortino ratio of 2.41 while incurring a maximum drawdown of -3.62\%, which is almost 3.5 times larger than the maximum drawdown incurred by the final unified momentum portfolio. When we compare the performance of \textsf{\modelname}-Fast, \textsf{\modelname}-Medium, and \textsf{\modelname}-Slow against TSMOM benchmark strategies, the results are less appealing. Across the board, \textsf{\modelname}-Fast, Medium, and Slow incur much higher maximum drawdowns compared to TSMOM strategies. Running a portfolio that incurs significantly larger drawdown risk without corresponding performance compensation may not be an appealing strategy for many portfolio managers.

Overall, the performance of the final unified momentum portfolio generated by \textsf{\modelnamecan} compared to the benchmark TSMOM strategies and \textsf{\modelname}-Fast, Medium, and Slow supports the claim that a portfolio capable of capitalizing on a spectrum of momentum opportunities results in a more robust portfolio, contributing to better risk-adjusted performance and minimizing maximum drawdown.

\begin{figure}[hp!]
\begin{subfigure}{1\textwidth}
  \centering
  % include first image
  \includegraphics[width=.6\linewidth]{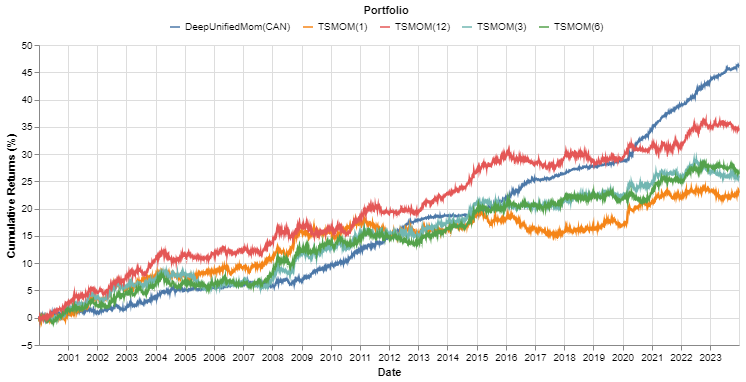}  
  \caption{\textsf{\modelnamecan} versus TSMOMs.}
  \label{fig:sub-first}
\end{subfigure}
\begin{subfigure}{1\textwidth}
  \centering
  % include second image
  \includegraphics[width=.6\linewidth]{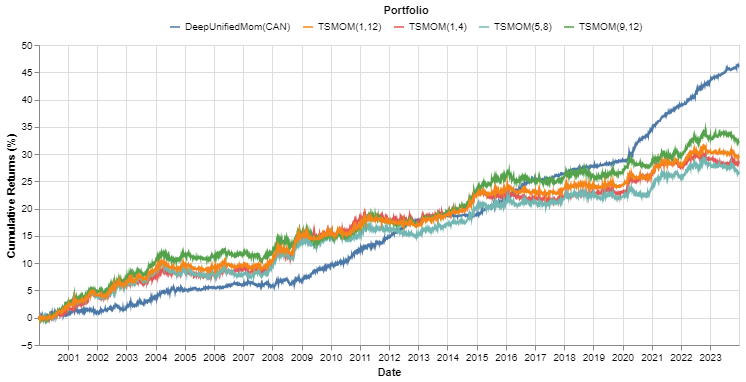}  
  \caption{\textsf{\modelnamecan} versus Equal Weight TSMOMs.}
  \label{fig:sub-second}
\end{subfigure}
\begin{subfigure}{1\textwidth}
  \centering
  % include third image
  \includegraphics[width=.6\linewidth]{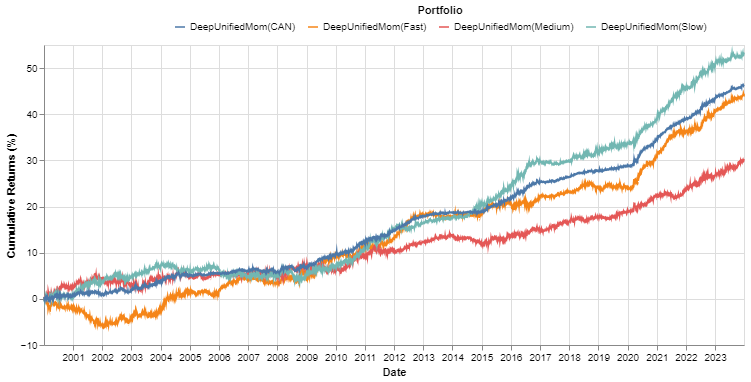}  
  \caption{\textsf{\modelnamecan} versus \textsf{\modelname} - Fast, Medium and Slow.}
  \label{fig:sub-third}
\end{subfigure}
\begin{subfigure}{1\textwidth}
  \centering
  % include fourth image
  \includegraphics[width=.6\linewidth]{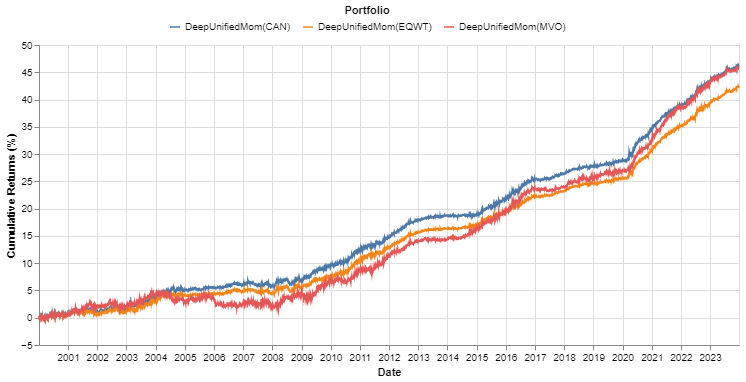}  
  \caption{\textsf{\modelnamecan} versus Baseline Portfolio Allocations: Equal Weight and MVO.}
  \label{fig:sub-fourth}
\end{subfigure}
\caption{Portfolios' Cumulative Returns (\%) from January 2000 to December 2023.}
\label{fig:fig}
\end{figure}

\begin{table}[htbp!]
\caption{Backtest results (net) for the period from January 2000 to December 2023, with transaction costs set at 3 basis points. The \textsf{\modelname} model results presented here were obtained using the Sharpe Ratio with a Soft Capping mechanism as the loss function during training. Max DD stands for maximum drawdown. }
\label{table:2}
\resizebox{\textwidth}{!}{%
\begin{tabular}{lccccc}
\toprule
\textbf{Portfolios} & \textbf{Ann. Return (\%)} & \textbf{Ann. vol (\%)} & \textbf{Sharpe} & \textbf{Sortino} & \textbf{Max DD (\%)}\\
 \midrule
TSMOM(1) & 1.06 & 1.44 & 0.73 & 1.09 & -4.60\\
TSMOM(3) & 1.16 & 1.46 & 0.80 & 1.16 & -3.37\\
TSMOM(6)  & 1.18 & 1.45 & 0.82 & 1.20 & -2.91\\
TSMOM(12)  & 1.48 & 1.47 & 1.01 & 1.46 & -3.28\\
\midrule
TSMOM(1,4)  & 1.27 & 1.27 & 1.00 & 1.49 & -2.05\\
TSMOM(5,8)  & 1.19 & 1.37 & 0.87 & 1.27 & -2.80\\
TSMOM(9,12) & 1.40 & 1.40 & 1.00 & 1.45 & -2.60\\
TSMOM(1,12) & 1.30 & 1.22 & 1.07 & 1.58 & -2.01\\
 \midrule
\textsf{\modelnamefast} & 1.85 & 1.39 & 1.34 & 2.06 & -6.32 \\
\textsf{\modelnamemedium} & 1.32 & 1.34 & 0.99 & 1.47 & -3.52 \\
\textsf{\modelnameslow} & \textbf{2.14} & 1.40 & 1.54 & 2.41 & -3.62 \\
 \midrule
\textsf{\modelnamecan} & 1.92 & 0.82 & \textbf{2.33} & \textbf{3.81} & -1.02 \\
\textsf{\modelnameeqwt} & 1.79 & \textbf{0.77} & 2.31 & 3.71 & \textbf{-0.99}\\
\textsf{\modelnamemvo} & 1.91 & 1.11 & 1.72 & 2.69 & -3.13\\
\bottomrule
\end{tabular}%
}
\end{table}

\begin{figure}[ht!]
  \centering
  \includegraphics[width=0.9\linewidth]{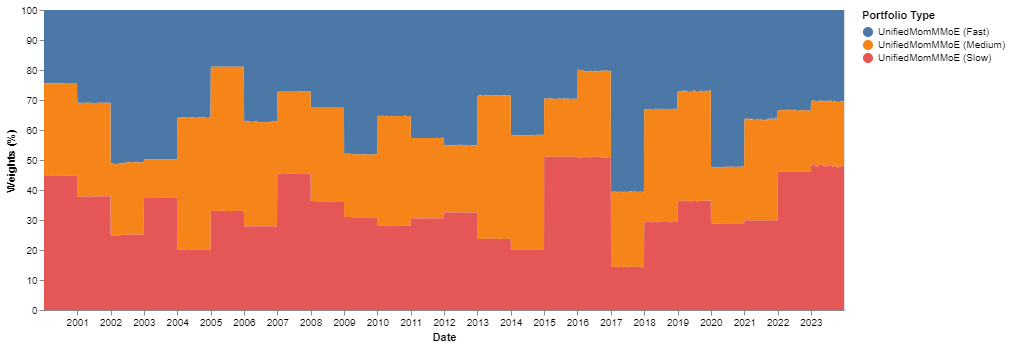}
  \caption{The figure illustrates the portfolio weight allocations determined by the Capital Allocation Task Specific Network within the \textsf{\modelname} framework across the Fast, Medium, and Slow portfolios from January 2000 to December 2023.}
  \label{fig:unified_mom_mmoe_capital_allocation}
\end{figure}

Figure \ref{fig:unified_mom_mmoe_capital_allocation} shows that the weights assigned by the \textsf{\modelnamecan} to the \textsf{\modelname} portfolios —Fast, Medium, and Slow— maintain remarkable consistency throughout the year. Moreover, the weight allocation strategy implemented by the \textsf{\modelnamecan} markedly diverges from traditional equal weighting approaches, demonstrating that \textsf{\modelnamecan} does not rely on a basic equal weighting scheme. However, \textsf{\modelnamecan} outperforms \textsf{\modelnameeqwt} by a narrow margin, achieving a Sharpe ratio of 2.33 compared to 2.31, and a Sortino ratio of 3.81 compared to 3.71. While \textsf{\modelnameeqwt} incurred a slightly smaller maximum drawdown of -0.99\%. Comparatively, both the \textsf{\modelnamecan} and \textsf{\modelnameeqwt} portfolios considerably outperform the \textsf{\modelnamemvo} approach in terms of risk-adjusted returns and maximum drawdown. The Sharpe ratio of \textsf{\modelnamemvo} portfolio is only 1.72, significantly lower than the 2.33 of the \textsf{\modelnamecan} portfolio. The maximum drawdown for the \textsf{\modelnamemvo} portfolio stands at -3.13\%, which is less favorable than the drawdowns experienced by the equal weighted TSMOM portfolios such as TSMOM(1,4), TSMOM(5,8), TSMOM(9,12), and TSMOM(1,12). These findings suggest that the \textsf{\modelnamemvo} approach may be suboptimal given that both \textsf{\modelnamecan} and \textsf{\modelnameeqwt} perform much better. While criticism of MVO is well-established, our results lend further support to these critiques, emphasizing the need for ongoing refinement and exploration of alternative strategies~\citep{michaud_2007, bailey_2012, jurczenko_2015, lopez_2016}. In conclusion, the performance analysis of \textsf{\modelnamecan} in comparison to benchmark strategies and \textsf{\modelname} portfolios —Fast, Medium, and Slow, underscores its effectiveness in portfolio management. By outperforming TSMOM benchmarks and task-specific networks in terms of risk-adjusted returns and drawdown management, \textsf{\modelnamecan} demonstrates its ability to capitalize on momentum opportunities across a spectrum of market conditions. This suggests that a unified approach to portfolio construction, such as \textsf{\modelnamecan}, leads to a more robust portfolio, contributing to improved risk-adjusted performance and minimized maximum drawdown. These findings highlight the potential of \textsf{\modelnamecan} as a practical and promising solution for investors seeking enhanced portfolio performance in dynamic market environments.

\begin{table}[htb]
\caption{Here are the backtest metrics (net) for the period from January 2000 to December 2023, with transaction costs set at 3 basis points. The \textsf{\modelname} model results presented here were obtained using the Sharpe Ratio with a Soft Capping mechanism as the loss function during training.}
\label{table:3}
\resizebox{\textwidth}{!}{%
\begin{tabular}{lccccc}
\toprule
\textbf{Portfolios} & \textbf{Ann. Return (\%)} & \textbf{Ann. vol (\%)} & \textbf{Sharpe} & \textbf{Sortino} & \textbf{Max DD (\%)}\\
 \midrule
 Sharpe Ratio with Soft Capping Mechanism (Threshold = 0.01)\\
 \midrule
\textsf{\modelnamefast} & 1.85 & 1.39 & 1.34 & 2.06 & -6.32 \\
\textsf{\modelnamemedium} & 1.32 & 1.34 & 0.99 & 1.47 & -3.52 \\
\textsf{\modelnameslow} & \textbf{2.14} & 1.40 & 1.54 & 2.41 & -3.62 \\
\textsf{\modelnamecan} & 1.92 & 0.82 & \textbf{2.33} & \textbf{3.81} & -1.02 \\
\textsf{\modelnameeqwt} & 1.79 & \textbf{0.77} & 2.31 & 3.71 & \textbf{-0.99}\\
\textsf{\modelnamemvo} & 1.91 & 1.11 & 1.72 & 2.69 & -3.13\\
\midrule
Sharpe Ratio \\
 \midrule
\textsf{\modelnamefast} & 1.73 & 1.42 & 1.22 & 1.87 & -6.33 \\
\textsf{\modelnamemedium} & 1.20 & 1.39 & 0.87 & 1.28 & -3.61 \\
\textsf{\modelnameslow} & \textbf{2.03} & 1.43 & 1.42 & 2.20 & -4.00 \\
\textsf{\modelnamecan} & 1.80 & 0.84 & \textbf{2.14} & \textbf{3.43} & \textbf{-1.10} \\
\textsf{\modelnameeqwt} & 1.67 & \textbf{0.79} & 2.11 & 3.33 & -1.21\\
\textsf{\modelnamemvo} & 1.72 & 1.17 & 1.47 & 2.26 & -3.64\\
\bottomrule
\end{tabular}%
}
\end{table}

Table \ref{table:3} reveals that when \textsf{\modelnamecan} is trained using the Sharpe Ratio with a Soft Capping Mechanism, the resulting portfolios consistently perform better than those trained using only the Sharpe Ratio. Specifically, \textsf{\modelnamecan} trained with the modified Sharpe Ratio achieved a Sharpe Ratio of 2.33 and a Sortino Ratio of 3.81, compared to 2.14 and 3.43, respectively, for those trained with the standard Sharpe Ratio. Additionally, the task-specific networks constructing the Fast, Medium, and Slow portfolios also outperform their counterparts. Overall, the results are promising, indicating that further research into improved Sharpe Ratio objective functions for training deep learning model is worthwhile.

\section{Conclusion}
\label{sec:conclusion}

The proposed \textsf{\modelname} framework represents a significant advancement in applying deep learning in portfolio management, adeptly addressing the limitations of traditional momentum strategies. At its core, \textsf{\modelname} leverages advanced deep learning, employing a multi-task learning approach and a multi-gate mixture of experts to construct unified momentum portfolios. Our extensive backtesting, spanning various asset classes such as equity indexes, bonds, currencies, and commodities, has consistently shown that \textsf{\modelname} surpasses benchmarks, maintaining its superior performance even after accounting for transaction costs. This highlights its ability to construct a final portfolio that accounts for a wide spectrum of momentum opportunities in the financial market in an end-to-end fashion. This new approach to using deep learning in investment strategies showcases the benefits of advanced computational techniques in financial decision-making and outcomes. The model was developed using Python and Pytorch; the framework is accessible for review and utilization, with its source code publicly available online\footnote{\url{https://github.com/joelowj/unified_mom_mmoe}}. Future research will focus on enhancing the \textsf{\modelname} framework by incorporating sparsity into the gating mechanism and utilizing more sophisticated deep learning architectures, such as the Transformer model, for time-series analysis.  Additionally, efforts will be made to integrate explainable AI techniques into the portfolio construction process, aiming to increase the transparency and interpretability of the \textsf{\modelname} framework. This progression will not only refine the model's performance but also bolster user trust and understanding of how AI-driven decisions are made within the portfolio management context.

\bibliography{sample}

\begin{thebibliography}{61}
\expandafter\ifx\csname natexlab\endcsname\relax\def\natexlab#1{#1}\fi
\providecommand{\url}[1]{\texttt{#1}}
\providecommand{\href}[2]{#2}
\providecommand{\path}[1]{#1}
\providecommand{\DOIprefix}{doi:}
\providecommand{\ArXivprefix}{arXiv:}
\providecommand{\URLprefix}{URL: }
\providecommand{\Pubmedprefix}{pmid:}
\providecommand{\doi}[1]{\href{http://dx.doi.org/#1}{\path{#1}}}
\providecommand{\Pubmed}[1]{\href{pmid:#1}{\path{#1}}}
\providecommand{\bibinfo}[2]{#2}
\ifx\xfnm\relax \def\xfnm[#1]{\unskip,\space#1}\fi
%Type = Article
\bibitem[{Asness et~al.(2014)Asness, Frazzini, Israel \& Moskowitz}]{asness_2014}
\bibinfo{author}{Asness, C.}, \bibinfo{author}{Frazzini, A.}, \bibinfo{author}{Israel, R.}, \& \bibinfo{author}{Moskowitz, T.} (\bibinfo{year}{2014}).
\newblock \bibinfo{title}{Fact, fiction, and momentum investing}.
\newblock {\it \bibinfo{journal}{The Journal of Portfolio Management}\/},  {\it \bibinfo{volume}{40}\/}, \bibinfo{pages}{75--92}.
%Type = Article
\bibitem[{Bailey \& Lopez~de Prado(2012)}]{bailey_2012}
\bibinfo{author}{Bailey, D.}, \& \bibinfo{author}{Lopez~de Prado, M.} (\bibinfo{year}{2012}).
\newblock \bibinfo{title}{The sharpe ratio efficient frontier}.
\newblock {\it \bibinfo{journal}{The Journal of Risk}\/},  {\it \bibinfo{volume}{15}\/}, \bibinfo{pages}{3--44}. \DOIprefix\doi{10.21314/JOR.2012.255}.
%Type = Article
\bibitem[{Bailey et~al.(2015)Bailey, Borwein, Lopez~de Prado \& Zhu}]{bailey2015probability}
\bibinfo{author}{Bailey, D.~H.}, \bibinfo{author}{Borwein, J.~M.}, \bibinfo{author}{Lopez~de Prado, M.}, \& \bibinfo{author}{Zhu, Q.~J.} (\bibinfo{year}{2015}).
\newblock \bibinfo{title}{The probability of backtest overfitting}.
\newblock {\it \bibinfo{journal}{Journal of Computational Finance}\/}, .
\newblock \bibinfo{note}{Forthcoming}.
%Type = Article
\bibitem[{Baltas \& Kosowski(2012)}]{baltas_2012}
\bibinfo{author}{Baltas, A.}, \& \bibinfo{author}{Kosowski, R.} (\bibinfo{year}{2012}).
\newblock \bibinfo{title}{Improving time-series momentum strategies: The role of trading signals and volatility estimators}.
\newblock {\it \bibinfo{journal}{SSRN Electronic Journal}\/}, .
%Type = Inbook
\bibitem[{Baltas \& Kosowski(2021)}]{baltas_2021}
\bibinfo{author}{Baltas, N.}, \& \bibinfo{author}{Kosowski, R.} (\bibinfo{year}{2021}).
\newblock \bibinfo{title}{Demystifying time-series momentum strategies: Volatility estimators, trading rules and pairwise correlations}.
\newblock In {\it \bibinfo{booktitle}{Market momentum theory and practice}\/} (p. \bibinfo{pages}{30–63}).
\newblock \bibinfo{publisher}{Wiley}.
%Type = Article
\bibitem[{Barroso \& Santa-Clara(2015)}]{barroso_2015}
\bibinfo{author}{Barroso, P.}, \& \bibinfo{author}{Santa-Clara, P.} (\bibinfo{year}{2015}).
\newblock \bibinfo{title}{Momentum has its moments}.
\newblock {\it \bibinfo{journal}{Journal of Financial Economics}\/},  {\it \bibinfo{volume}{116}\/}, \bibinfo{pages}{111--120}.
%Type = Article
\bibitem[{Baxter(2000)}]{baxter_2000}
\bibinfo{author}{Baxter, J.} (\bibinfo{year}{2000}).
\newblock \bibinfo{title}{A model of inductive bias learning}.
\newblock {\it \bibinfo{journal}{J. Artif. Int. Res.}\/},  {\it \bibinfo{volume}{12}\/}, \bibinfo{pages}{149–198}.
%Type = Article
\bibitem[{Caruana(1997)}]{caruana_1997}
\bibinfo{author}{Caruana, R.} (\bibinfo{year}{1997}).
\newblock \bibinfo{title}{Multitask learning}.
\newblock {\it \bibinfo{journal}{Machine Learning}\/},  {\it \bibinfo{volume}{28}\/}, \bibinfo{pages}{41--75}.
%Type = Article
\bibitem[{Daniel \& Moskowitz(2016)}]{daniel_2016}
\bibinfo{author}{Daniel, K.}, \& \bibinfo{author}{Moskowitz, T.~J.} (\bibinfo{year}{2016}).
\newblock \bibinfo{title}{Momentum crashes}.
\newblock {\it \bibinfo{journal}{Journal of Financial Economics}\/},  {\it \bibinfo{volume}{122}\/}, \bibinfo{pages}{221--247}.
%Type = Article
\bibitem[{Duan(2023)}]{duan_2023}
\bibinfo{author}{Duan, S.} (\bibinfo{year}{2023}).
\newblock \bibinfo{title}{Performance of time-series momentum strategy: Us evidence}.
\newblock {\it \bibinfo{journal}{Advances in Economics, Management and Political Sciences}\/},  {\it \bibinfo{volume}{35}\/}, \bibinfo{pages}{45--54}. \DOIprefix\doi{10.54254/2754-1169/35/20231722}.
%Type = Article
\bibitem[{Fedus et~al.(2022)Fedus, Zoph \& Shazeer}]{fedus_2022}
\bibinfo{author}{Fedus, W.}, \bibinfo{author}{Zoph, B.}, \& \bibinfo{author}{Shazeer, N.} (\bibinfo{year}{2022}).
\newblock \bibinfo{title}{Switch transformers: scaling to trillion parameter models with simple and efficient sparsity}.
\newblock {\it \bibinfo{journal}{J. Mach. Learn. Res.}\/},  {\it \bibinfo{volume}{23}\/}.
%Type = Misc
\bibitem[{Gale et~al.(2022)Gale, Narayanan, Young \& Zaharia}]{gale_2022}
\bibinfo{author}{Gale, T.}, \bibinfo{author}{Narayanan, D.}, \bibinfo{author}{Young, C.}, \& \bibinfo{author}{Zaharia, M.} (\bibinfo{year}{2022}).
\newblock \bibinfo{title}{Megablocks: Efficient sparse training with mixture-of-experts}.
\newblock \href{http://arxiv.org/abs/2211.15841}{\tt arXiv:2211.15841}.
%Type = Article
\bibitem[{Garman \& Klass(1980)}]{garman_1980}
\bibinfo{author}{Garman, M.~B.}, \& \bibinfo{author}{Klass, M.~J.} (\bibinfo{year}{1980}).
\newblock \bibinfo{title}{On the estimation of security price volatilities from historical data}.
\newblock {\it \bibinfo{journal}{The Journal of Business}\/},  {\it \bibinfo{volume}{53}\/}, \bibinfo{pages}{67--78}.
%Type = Article
\bibitem[{Georgopoulou \& Wang(2016)}]{georgopoulou_wang_2016}
\bibinfo{author}{Georgopoulou, A.~A.}, \& \bibinfo{author}{Wang, G.~J.} (\bibinfo{year}{2016}).
\newblock \bibinfo{title}{The trend is your friend: Time-series momentum strategies across equity and commodity markets}.
\newblock {\it \bibinfo{journal}{SSRN Electronic Journal}\/}, .
%Type = Inproceedings
\bibitem[{Ghosn \& Bengio(1996)}]{ghosn_1996}
\bibinfo{author}{Ghosn, J.}, \& \bibinfo{author}{Bengio, Y.} (\bibinfo{year}{1996}).
\newblock \bibinfo{title}{Multi-task learning for stock selection}.
\newblock In \bibinfo{editor}{M.~Mozer}, \bibinfo{editor}{M.~Jordan}, \& \bibinfo{editor}{T.~Petsche} (Eds.), {\it \bibinfo{booktitle}{Advances in Neural Information Processing Systems}\/}.
\newblock \bibinfo{publisher}{MIT Press} volume~\bibinfo{volume}{9}.
\newblock \URLprefix \url{https://proceedings.neurips.cc/paper_files/paper/1996/file/1d72310edc006dadf2190caad5802983-Paper.pdf}.
%Type = Article
\bibitem[{Goulding et~al.(2023)Goulding, Harvey \& Mazzoleni}]{goulding_2023}
\bibinfo{author}{Goulding, C.~L.}, \bibinfo{author}{Harvey, C.~R.}, \& \bibinfo{author}{Mazzoleni, M.} (\bibinfo{year}{2023}).
\newblock \bibinfo{title}{Momentum turning points}.
\newblock {\it \bibinfo{journal}{SSRN Electronic Journal}\/}, .
%Type = Article
\bibitem[{Harvey et~al.(2020)Harvey, Hoyle, Rattray \& van Hemert}]{cambell_2020}
\bibinfo{author}{Harvey, C.~R.}, \bibinfo{author}{Hoyle, E.}, \bibinfo{author}{Rattray, S.}, \& \bibinfo{author}{van Hemert, O.} (\bibinfo{year}{2020}).
\newblock \bibinfo{title}{Strategic risk management: Out-of-sample evidence from the covid-19 equity selloff}.
\newblock {\it \bibinfo{journal}{SSRN Electronic Journal}\/}, .
%Type = Article
\bibitem[{Harvey \& Liu(2014)}]{harvey2014evaluating}
\bibinfo{author}{Harvey, C.~R.}, \& \bibinfo{author}{Liu, Y.} (\bibinfo{year}{2014}).
\newblock \bibinfo{title}{Evaluating trading strategies}.
\newblock {\it \bibinfo{journal}{The Journal of Portfolio Management}\/},  {\it \bibinfo{volume}{40}\/}, \bibinfo{pages}{108--118}.
%Type = Inproceedings
\bibitem[{He et~al.(2022)He, Zhai, Antunes, Wang, Luo, Shi \& Li}]{he_jiaao_2022}
\bibinfo{author}{He, J.}, \bibinfo{author}{Zhai, J.}, \bibinfo{author}{Antunes, T.}, \bibinfo{author}{Wang, H.}, \bibinfo{author}{Luo, F.}, \bibinfo{author}{Shi, S.}, \& \bibinfo{author}{Li, Q.} (\bibinfo{year}{2022}).
\newblock \bibinfo{title}{Fastermoe: modeling and optimizing training of large-scale dynamic pre-trained models}.
\newblock In {\it \bibinfo{booktitle}{Proceedings of the 27th ACM SIGPLAN Symposium on Principles and Practice of Parallel Programming}\/} PPoPP '22 (p. \bibinfo{pages}{120–134}).
\newblock \bibinfo{address}{New York, NY, USA}: \bibinfo{publisher}{Association for Computing Machinery}.
\newblock \URLprefix \url{https://doi.org/10.1145/3503221.3508418}. \DOIprefix\doi{10.1145/3503221.3508418}.
%Type = Article
\bibitem[{Hochreiter \& Schmidhuber(1997)}]{schmidhuber_1997}
\bibinfo{author}{Hochreiter, S.}, \& \bibinfo{author}{Schmidhuber, J.} (\bibinfo{year}{1997}).
\newblock \bibinfo{title}{{Long Short-Term Memory}}.
\newblock {\it \bibinfo{journal}{Neural Computation}\/},  {\it \bibinfo{volume}{9}\/}, \bibinfo{pages}{1735--1780}. \DOIprefix\doi{10.1162/neco.1997.9.8.1735}.
%Type = Article
\bibitem[{Hurst et~al.(2017)Hurst, Ooi \& Pedersen}]{hurst_2017}
\bibinfo{author}{Hurst, B.}, \bibinfo{author}{Ooi, Y.~H.}, \& \bibinfo{author}{Pedersen, L.~H.} (\bibinfo{year}{2017}).
\newblock \bibinfo{title}{A century of evidence on trend-following investing}.
\newblock {\it \bibinfo{journal}{The Journal of Portfolio Management}\/},  {\it \bibinfo{volume}{44}\/}, \bibinfo{pages}{15--29}.
%Type = Article
\bibitem[{Israel et~al.(2020)Israel, Kelly \& Moskowitz}]{israel2020can}
\bibinfo{author}{Israel, R.}, \bibinfo{author}{Kelly, B.}, \& \bibinfo{author}{Moskowitz, T.} (\bibinfo{year}{2020}).
\newblock \bibinfo{title}{Can machines “learn” finance?}
\newblock {\it \bibinfo{journal}{Journal of Investment Management}\/},  {\it \bibinfo{volume}{18}\/}. \URLprefix \url{https://joim.com/article/can-machines-learn-finance/}.
%Type = Article
\bibitem[{Jacobs et~al.(1991)Jacobs, Jordan, Nowlan \& Hinton}]{jacobs_1991}
\bibinfo{author}{Jacobs, R.~A.}, \bibinfo{author}{Jordan, M.~I.}, \bibinfo{author}{Nowlan, S.~J.}, \& \bibinfo{author}{Hinton, G.~E.} (\bibinfo{year}{1991}).
\newblock \bibinfo{title}{Adaptive mixtures of local experts}.
\newblock {\it \bibinfo{journal}{Neural Computation}\/},  {\it \bibinfo{volume}{3}\/}, \bibinfo{pages}{79--87}. \DOIprefix\doi{10.1162/neco.1991.3.1.79}.
%Type = Article
\bibitem[{Jegadeesh \& Titman(1993)}]{jegadeesh_titman_1993}
\bibinfo{author}{Jegadeesh, N.}, \& \bibinfo{author}{Titman, S.} (\bibinfo{year}{1993}).
\newblock \bibinfo{title}{Returns to buying winners and selling losers: Implications for stock market efficiency}.
\newblock {\it \bibinfo{journal}{The Journal of Finance}\/},  {\it \bibinfo{volume}{48}\/}, \bibinfo{pages}{65--91}.
%Type = Article
\bibitem[{Jegadeesh \& Titman(2001)}]{jegadeesh_titman_2001}
\bibinfo{author}{Jegadeesh, N.}, \& \bibinfo{author}{Titman, S.} (\bibinfo{year}{2001}).
\newblock \bibinfo{title}{Profitability of momentum strategies: An evaluation of alternative explanations}.
\newblock {\it \bibinfo{journal}{The Journal of Finance}\/},  {\it \bibinfo{volume}{56}\/}, \bibinfo{pages}{699--720}.
%Type = Article
\bibitem[{Jurczenko \& Teiletche(2015)}]{jurczenko_2015}
\bibinfo{author}{Jurczenko, E.}, \& \bibinfo{author}{Teiletche, J.} (\bibinfo{year}{2015}).
\newblock \bibinfo{title}{Active risk-based investing}.
\newblock {\it \bibinfo{journal}{SSRN Electronic Journal}\/}, . \DOIprefix\doi{10.2139/ssrn.2592904}.
%Type = Article
\bibitem[{Kang et~al.(2022)Kang, Chen, Jia, Wei, Deng \& Qian}]{yanzhe_2022}
\bibinfo{author}{Kang, Y.}, \bibinfo{author}{Chen, L.}, \bibinfo{author}{Jia, N.}, \bibinfo{author}{Wei, W.}, \bibinfo{author}{Deng, J.}, \& \bibinfo{author}{Qian, H.} (\bibinfo{year}{2022}).
\newblock \bibinfo{title}{A cwgan-gp-based multi-task learning model for consumer credit scoring}.
\newblock {\it \bibinfo{journal}{Expert Systems with Applications}\/},  {\it \bibinfo{volume}{206}\/}, \bibinfo{pages}{117650}. \URLprefix \url{https://www.sciencedirect.com/science/article/pii/S0957417422009538}. \DOIprefix\doi{https://doi.org/10.1016/j.eswa.2022.117650}.
%Type = Article
\bibitem[{Levine \& Pedersen(2016)}]{levine_2016}
\bibinfo{author}{Levine, A.}, \& \bibinfo{author}{Pedersen, L.~H.} (\bibinfo{year}{2016}).
\newblock \bibinfo{title}{Which trend is your friend?}
\newblock {\it \bibinfo{journal}{Financial Analysts Journal}\/},  {\it \bibinfo{volume}{72}\/}, \bibinfo{pages}{51--66}.
%Type = Article
\bibitem[{Liebel \& K{\"{o}}rner(2018)}]{lukas_2018}
\bibinfo{author}{Liebel, L.}, \& \bibinfo{author}{K{\"{o}}rner, M.} (\bibinfo{year}{2018}).
\newblock \bibinfo{title}{Auxiliary tasks in multi-task learning}.
\newblock {\it \bibinfo{journal}{CoRR}\/},  {\it \bibinfo{volume}{abs/1805.06334}\/}. \href{http://arxiv.org/abs/1805.06334}{\tt arXiv:1805.06334}.
%Type = Article
\bibitem[{Lim et~al.(2019)Lim, Zohren \& Roberts}]{bryan_2019}
\bibinfo{author}{Lim, B.}, \bibinfo{author}{Zohren, S.}, \& \bibinfo{author}{Roberts, S.} (\bibinfo{year}{2019}).
\newblock \bibinfo{title}{Enhancing time series momentum strategies using deep neural networks}.
\newblock {\it \bibinfo{journal}{arXiv preprint arXiv:1904.04912}\/}, . \DOIprefix\doi{10.48550/ARXIV.1904.04912}.
%Type = Article
\bibitem[{Lin et~al.(2022)Lin, Chen, Sang \& Huang}]{lin_2022}
\bibinfo{author}{Lin, Y.-C.}, \bibinfo{author}{Chen, C.-T.}, \bibinfo{author}{Sang, C.-Y.}, \& \bibinfo{author}{Huang, S.-H.} (\bibinfo{year}{2022}).
\newblock \bibinfo{title}{Multiagent-based deep reinforcement learning for risk-shifting portfolio management}.
\newblock {\it \bibinfo{journal}{Applied Soft Computing}\/},  {\it \bibinfo{volume}{123}\/}, \bibinfo{pages}{108894}. \URLprefix \url{https://www.sciencedirect.com/science/article/pii/S1568494622002678}. \DOIprefix\doi{https://doi.org/10.1016/j.asoc.2022.108894}.
%Type = Inproceedings
\bibitem[{Ma et~al.(2018)Ma, Zhao, Yi, Chen, Hong \& Chi}]{ma_2018}
\bibinfo{author}{Ma, J.}, \bibinfo{author}{Zhao, Z.}, \bibinfo{author}{Yi, X.}, \bibinfo{author}{Chen, J.}, \bibinfo{author}{Hong, L.}, \& \bibinfo{author}{Chi, E.~H.} (\bibinfo{year}{2018}).
\newblock \bibinfo{title}{Modeling task relationships in multi-task learning with multi-gate mixture-of-experts}.
\newblock In {\it \bibinfo{booktitle}{Proceedings of the 24th ACM SIGKDD International Conference on Knowledge Discovery \& Data Mining}\/} KDD '18 (p. \bibinfo{pages}{1930–1939}).
\newblock \bibinfo{address}{New York, NY, USA}: \bibinfo{publisher}{Association for Computing Machinery}.
\newblock \URLprefix \url{https://doi.org/10.1145/3219819.3220007}. \DOIprefix\doi{10.1145/3219819.3220007}.
%Type = Article
\bibitem[{Malitskaia(2020)}]{yulia_2020}
\bibinfo{author}{Malitskaia, Y.} (\bibinfo{year}{2020}).
\newblock \bibinfo{title}{Uncovering momentum}.
\newblock {\it \bibinfo{journal}{SSRN Electronic Journal}\/}, .
%Type = Article
\bibitem[{Markowitz(1952)}]{harry_1952}
\bibinfo{author}{Markowitz, H.} (\bibinfo{year}{1952}).
\newblock \bibinfo{title}{Portfolio selection}.
\newblock {\it \bibinfo{journal}{The Journal of Finance}\/},  {\it \bibinfo{volume}{7}\/}, \bibinfo{pages}{77--91}.
%Type = Article
\bibitem[{Michaud \& Michaud(2007)}]{michaud_2007}
\bibinfo{author}{Michaud, R.}, \& \bibinfo{author}{Michaud, R.} (\bibinfo{year}{2007}).
\newblock \bibinfo{title}{Estimation error and portfolio optimization: A resampling solution}.
\newblock {\it \bibinfo{journal}{Journal of Investment Management}\/},  {\it \bibinfo{volume}{Vol. 6}\/}, \bibinfo{pages}{pp. 8 -- 28}.
%Type = Article
\bibitem[{Min et~al.(2021)Min, Dong, Liu \& Gong}]{min_2021}
\bibinfo{author}{Min, L.}, \bibinfo{author}{Dong, J.}, \bibinfo{author}{Liu, J.}, \& \bibinfo{author}{Gong, X.} (\bibinfo{year}{2021}).
\newblock \bibinfo{title}{Robust mean-risk portfolio optimization using machine learning-based trade-off parameter}.
\newblock {\it \bibinfo{journal}{Applied Soft Computing}\/},  {\it \bibinfo{volume}{113}\/}, \bibinfo{pages}{107948}. \URLprefix \url{https://www.sciencedirect.com/science/article/pii/S156849462100870X}. \DOIprefix\doi{https://doi.org/10.1016/j.asoc.2021.107948}.
%Type = Article
\bibitem[{Moskowitz et~al.(2012)Moskowitz, Ooi \& Pedersen}]{moskowitz_2012}
\bibinfo{author}{Moskowitz, T.~J.}, \bibinfo{author}{Ooi, Y.~H.}, \& \bibinfo{author}{Pedersen, L.~H.} (\bibinfo{year}{2012}).
\newblock \bibinfo{title}{Time series momentum}.
\newblock {\it \bibinfo{journal}{Journal of Financial Economics}\/},  {\it \bibinfo{volume}{104}\/}, \bibinfo{pages}{228--250}.
%Type = Article
\bibitem[{Nilsson(2015)}]{nilsson_2015}
\bibinfo{author}{Nilsson, L.} (\bibinfo{year}{2015}).
\newblock \bibinfo{title}{Trend following - expected returns}.
\newblock {\it \bibinfo{journal}{SSRN Electronic Journal}\/}, .
%Type = Article
\bibitem[{Ong \& Herremans(2023)}]{ong_2023}
\bibinfo{author}{Ong, J.}, \& \bibinfo{author}{Herremans, D.} (\bibinfo{year}{2023}).
\newblock \bibinfo{title}{Constructing time-series momentum portfolios with deep multi-task learning}.
\newblock {\it \bibinfo{journal}{Expert Systems with Applications}\/},  {\it \bibinfo{volume}{230}\/}, \bibinfo{pages}{120587}.
%Type = Article
\bibitem[{Ozbayoglu et~al.(2020)Ozbayoglu, Gudelek \& Sezer}]{ozbayoglu_2020}
\bibinfo{author}{Ozbayoglu, A.~M.}, \bibinfo{author}{Gudelek, M.~U.}, \& \bibinfo{author}{Sezer, O.~B.} (\bibinfo{year}{2020}).
\newblock \bibinfo{title}{Deep learning for financial applications : A survey}.
\newblock {\it \bibinfo{journal}{Applied Soft Computing}\/},  {\it \bibinfo{volume}{93}\/}, \bibinfo{pages}{106384}. \URLprefix \url{https://www.sciencedirect.com/science/article/pii/S1568494620303240}. \DOIprefix\doi{https://doi.org/10.1016/j.asoc.2020.106384}.
%Type = Article
\bibitem[{Parkinson(1980)}]{parkinson_1980}
\bibinfo{author}{Parkinson, M.} (\bibinfo{year}{1980}).
\newblock \bibinfo{title}{The extreme value method for estimating the variance of the rate of return}.
\newblock {\it \bibinfo{journal}{The Journal of Business}\/},  {\it \bibinfo{volume}{53}\/}, \bibinfo{pages}{61--65}.
%Type = Article
\bibitem[{Lopez~de Prado(2016)}]{lopez_2016}
\bibinfo{author}{Lopez~de Prado, M.} (\bibinfo{year}{2016}).
\newblock \bibinfo{title}{Building diversified portfolios that outperform out of sample:}.
\newblock {\it \bibinfo{journal}{The Journal of Portfolio Management}\/},  {\it \bibinfo{volume}{42}\/}, \bibinfo{pages}{59--69}. \DOIprefix\doi{10.3905/jpm.2016.42.4.059}.
%Type = Book
\bibitem[{L{\'o}pez~de Prado(2018)}]{lopez2018advances}
\bibinfo{author}{L{\'o}pez~de Prado, M.} (\bibinfo{year}{2018}).
\newblock {\it \bibinfo{title}{Advances in Financial Machine Learning}\/}.
\newblock \bibinfo{publisher}{Wiley}.
%Type = Misc
\bibitem[{Pu et~al.(2023)Pu, Roberts, Dong \& Zohren}]{xingyue_2023}
\bibinfo{author}{Pu, X.}, \bibinfo{author}{Roberts, S.}, \bibinfo{author}{Dong, X.}, \& \bibinfo{author}{Zohren, S.} (\bibinfo{year}{2023}).
\newblock \bibinfo{title}{Network momentum across asset classes}.
\newblock \href{http://arxiv.org/abs/2308.11294}{\tt arXiv:2308.11294}.
%Type = Article
\bibitem[{Rogers \& Satchell(1991)}]{rogers_1991}
\bibinfo{author}{Rogers, L. C.~G.}, \& \bibinfo{author}{Satchell, S.~E.} (\bibinfo{year}{1991}).
\newblock \bibinfo{title}{Estimating variance from high, low and closing prices}.
\newblock {\it \bibinfo{journal}{The Annals of Applied Probability}\/},  {\it \bibinfo{volume}{1}\/}, \bibinfo{pages}{504--512}.
%Type = Article
\bibitem[{Ruder(2017)}]{sebastian_2017}
\bibinfo{author}{Ruder, S.} (\bibinfo{year}{2017}).
\newblock \bibinfo{title}{An overview of multi-task learning in deep neural networks}.
\newblock {\it \bibinfo{journal}{CoRR}\/},  {\it \bibinfo{volume}{abs/1706.05098}\/}. \href{http://arxiv.org/abs/1706.05098}{\tt arXiv:1706.05098}.
%Type = Article
\bibitem[{Sebastian~Thrun(1998)}]{thrun_1998}
\bibinfo{author}{Sebastian~Thrun, L.~P.} (\bibinfo{year}{1998}).
\newblock \bibinfo{title}{Learning to learn: Introduction and overview}.
\newblock {\it \bibinfo{journal}{Learning to Learn}\/},  (pp. \bibinfo{pages}{3--17}).
%Type = Article
\bibitem[{Sharpe(1994)}]{sharpe_1994}
\bibinfo{author}{Sharpe, W.~F.} (\bibinfo{year}{1994}).
\newblock \bibinfo{title}{The sharpe ratio}.
\newblock {\it \bibinfo{journal}{The Journal of Portfolio Management}\/},  {\it \bibinfo{volume}{21}\/}, \bibinfo{pages}{49--58}.
%Type = Misc
\bibitem[{Shen et~al.(2023)Shen, Hou, Zhou, Du, Longpre, Wei, Chung, Zoph, Fedus, Chen, Vu, Wu, Chen, Webson, Li, Zhao, Yu, Keutzer, Darrell \& Zhou}]{sheng_2023}
\bibinfo{author}{Shen, S.}, \bibinfo{author}{Hou, L.}, \bibinfo{author}{Zhou, Y.}, \bibinfo{author}{Du, N.}, \bibinfo{author}{Longpre, S.}, \bibinfo{author}{Wei, J.}, \bibinfo{author}{Chung, H.~W.}, \bibinfo{author}{Zoph, B.}, \bibinfo{author}{Fedus, W.}, \bibinfo{author}{Chen, X.}, \bibinfo{author}{Vu, T.}, \bibinfo{author}{Wu, Y.}, \bibinfo{author}{Chen, W.}, \bibinfo{author}{Webson, A.}, \bibinfo{author}{Li, Y.}, \bibinfo{author}{Zhao, V.}, \bibinfo{author}{Yu, H.}, \bibinfo{author}{Keutzer, K.}, \bibinfo{author}{Darrell, T.}, \& \bibinfo{author}{Zhou, D.} (\bibinfo{year}{2023}).
\newblock \bibinfo{title}{Mixture-of-experts meets instruction tuning:a winning combination for large language models}.
\newblock \href{http://arxiv.org/abs/2305.14705}{\tt arXiv:2305.14705}.
%Type = Article
\bibitem[{Tzotchev(2018)}]{tzotchev_2018}
\bibinfo{author}{Tzotchev, D.} (\bibinfo{year}{2018}).
\newblock \bibinfo{title}{Designing robust trend-following system: Behind the scenes of trend-following}.
\newblock {\it \bibinfo{journal}{SSRN Electronic Journal}\/}, .
%Type = Inproceedings
\bibitem[{Vaswani et~al.(2017)Vaswani, Shazeer, Parmar, Uszkoreit, Jones, Gomez, Kaiser \& Polosukhin}]{vaswani_2017}
\bibinfo{author}{Vaswani, A.}, \bibinfo{author}{Shazeer, N.}, \bibinfo{author}{Parmar, N.}, \bibinfo{author}{Uszkoreit, J.}, \bibinfo{author}{Jones, L.}, \bibinfo{author}{Gomez, A.~N.}, \bibinfo{author}{Kaiser, L.~u.}, \& \bibinfo{author}{Polosukhin, I.} (\bibinfo{year}{2017}).
\newblock \bibinfo{title}{Attention is all you need}.
\newblock In \bibinfo{editor}{I.~Guyon}, \bibinfo{editor}{U.~V. Luxburg}, \bibinfo{editor}{S.~Bengio}, \bibinfo{editor}{H.~Wallach}, \bibinfo{editor}{R.~Fergus}, \bibinfo{editor}{S.~Vishwanathan}, \& \bibinfo{editor}{R.~Garnett} (Eds.), {\it \bibinfo{booktitle}{Advances in Neural Information Processing Systems}\/}.
\newblock \bibinfo{publisher}{Curran Associates, Inc.} volume~\bibinfo{volume}{30}.
%Type = Article
\bibitem[{Vitells \& Gross(2011)}]{vitells2011estimating}
\bibinfo{author}{Vitells, O.}, \& \bibinfo{author}{Gross, E.} (\bibinfo{year}{2011}).
\newblock \bibinfo{title}{Estimating the significance of a signal in a multi-dimensional search}.
\newblock {\it \bibinfo{journal}{Astroparticle Physics}\/},  {\it \bibinfo{volume}{35}\/}, \bibinfo{pages}{230--234}. \DOIprefix\doi{10.1016/j.astropartphys.2011.08.005}.
%Type = Article
\bibitem[{Wang et~al.(2022)Wang, Zhang \& Luo}]{wang_2022}
\bibinfo{author}{Wang, J.}, \bibinfo{author}{Zhang, H.}, \& \bibinfo{author}{Luo, H.} (\bibinfo{year}{2022}).
\newblock \bibinfo{title}{Research on the construction of stock portfolios based on multiobjective water cycle algorithm and kmv algorithm}.
\newblock {\it \bibinfo{journal}{Applied Soft Computing}\/},  {\it \bibinfo{volume}{115}\/}, \bibinfo{pages}{108186}. \URLprefix \url{https://www.sciencedirect.com/science/article/pii/S1568494621010346}. \DOIprefix\doi{https://doi.org/10.1016/j.asoc.2021.108186}.
%Type = Misc
\bibitem[{Wood et~al.(2022)Wood, Giegerich, Roberts \& Zohren}]{wood_giegerich_roberts_zohren_2022}
\bibinfo{author}{Wood, K.}, \bibinfo{author}{Giegerich, S.}, \bibinfo{author}{Roberts, S.}, \& \bibinfo{author}{Zohren, S.} (\bibinfo{year}{2022}).
\newblock \bibinfo{title}{Trading with the momentum transformer: An intelligent and interpretable architecture}.
\newblock \URLprefix \url{https://arxiv.org/abs/2112.08534}.
%Type = Article
\bibitem[{Yang \& Zhang(2000)}]{yang_2000}
\bibinfo{author}{Yang, D.}, \& \bibinfo{author}{Zhang, Q.} (\bibinfo{year}{2000}).
\newblock \bibinfo{title}{Drift‐independent volatility estimation based on high, low, open, and close prices}.
\newblock {\it \bibinfo{journal}{The Journal of Business}\/},  {\it \bibinfo{volume}{73}\/}, \bibinfo{pages}{477--492}.
%Type = Inbook
\bibitem[{Yu et~al.(2022)Yu, Wynter \& Lim}]{yu_2022}
\bibinfo{author}{Yu, P.}, \bibinfo{author}{Wynter, L.}, \& \bibinfo{author}{Lim, S.~H.} (\bibinfo{year}{2022}).
\newblock \bibinfo{title}{Federated reinforcement learning for portfolio management}.
\newblock In \bibinfo{editor}{H.~Ludwig}, \& \bibinfo{editor}{N.~Baracaldo} (Eds.), {\it \bibinfo{booktitle}{Federated Learning: A Comprehensive Overview of Methods and Applications}\/} (pp. \bibinfo{pages}{467--482}).
\newblock \bibinfo{address}{Cham}: \bibinfo{publisher}{Springer International Publishing}.
\newblock \URLprefix \url{https://doi.org/10.1007/978-3-030-96896-0_21}. \DOIprefix\doi{10.1007/978-3-030-96896-0_21}.
%Type = Article
\bibitem[{Yuan et~al.(2023)Yuan, Ma, Wang, Zhang \& Li}]{chenxun_2023}
\bibinfo{author}{Yuan, C.}, \bibinfo{author}{Ma, X.}, \bibinfo{author}{Wang, H.}, \bibinfo{author}{Zhang, C.}, \& \bibinfo{author}{Li, X.} (\bibinfo{year}{2023}).
\newblock \bibinfo{title}{Covid19-mlsf: A multi-task learning-based stock market forecasting framework during the covid-19 pandemic}.
\newblock {\it \bibinfo{journal}{Expert Systems with Applications}\/},  {\it \bibinfo{volume}{217}\/}, \bibinfo{pages}{119549}. \URLprefix \url{https://www.sciencedirect.com/science/article/pii/S0957417423000507}. \DOIprefix\doi{https://doi.org/10.1016/j.eswa.2023.119549}.
%Type = Article
\bibitem[{Zambrano \& Rizzolo(2022)}]{zambrano_2022}
\bibinfo{author}{Zambrano, E.~A.}, \& \bibinfo{author}{Rizzolo, C.} (\bibinfo{year}{2022}).
\newblock \bibinfo{title}{Long-only multi-asset momentum: Searching for absolute returns}.
\newblock {\it \bibinfo{journal}{SSRN Electronic Journal}\/}, .
%Type = Misc
\bibitem[{Zhang et~al.(2020)Zhang, Zohren \& Roberts}]{zhang_zohren_roberts_2020}
\bibinfo{author}{Zhang, Z.}, \bibinfo{author}{Zohren, S.}, \& \bibinfo{author}{Roberts, S.} (\bibinfo{year}{2020}).
\newblock \bibinfo{title}{Deep learning for portfolio optimisation}.
\newblock \URLprefix \url{https://papers.ssrn.com/sol3/papers.cfm?abstract_id=3613600}.
%Type = Article
\bibitem[{Zhao \& Yang(2023)}]{zhao_2023}
\bibinfo{author}{Zhao, Y.}, \& \bibinfo{author}{Yang, G.} (\bibinfo{year}{2023}).
\newblock \bibinfo{title}{Deep learning-based integrated framework for stock price movement prediction}.
\newblock {\it \bibinfo{journal}{Applied Soft Computing}\/},  {\it \bibinfo{volume}{133}\/}, \bibinfo{pages}{109921}. \URLprefix \url{https://www.sciencedirect.com/science/article/pii/S156849462200970X}. \DOIprefix\doi{https://doi.org/10.1016/j.asoc.2022.109921}.
%Type = Misc
\bibitem[{Zoph et~al.(2022)Zoph, Bello, Kumar, Du, Huang, Dean, Shazeer \& Fedus}]{zoph_2022}
\bibinfo{author}{Zoph, B.}, \bibinfo{author}{Bello, I.}, \bibinfo{author}{Kumar, S.}, \bibinfo{author}{Du, N.}, \bibinfo{author}{Huang, Y.}, \bibinfo{author}{Dean, J.}, \bibinfo{author}{Shazeer, N.}, \& \bibinfo{author}{Fedus, W.} (\bibinfo{year}{2022}).
\newblock \bibinfo{title}{St-moe: Designing stable and transferable sparse expert models}.
\newblock \href{http://arxiv.org/abs/2202.08906}{\tt arXiv:2202.08906}.

\end{thebibliography}

\end{document}